\documentclass{aipproc}
\layoutstyle{8x11single}
\usepackage{amsmath}
\usepackage{amssymb}

\usepackage{axodraw}
\newcommand{\ie}{{\it i.e.}}
\newcommand{\eg}{{\it e.g.}}
\newcommand{\cf}{{\it cf.}}
\newcommand{\rhs}{{\it rhs.\ }}

\newcommand{\gev}{{\rm GeV}}

\newcommand{\kvec}{\vect{k}}
\newcommand{\pvec}{\vect{p}}
\newcommand{\Pvec}{\vect{P}}
\newcommand{\qvec}{\vect{q}}

\newcommand{\xvec}{\vect{x}}

\newcommand{\bs}[1]{\boldsymbol{#1}}
\newcommand{\vect}[1]{{\bf{#1}}}

\newcommand{\as}{\alpha_s}

\newcommand{\lsim}{\buildrel < \over {_\sim}}

\newcommand{\order}[1]{${\cal O}\left(#1 \right)$}
\newcommand{\morder}[1]{{\cal O}\left(#1 \right)}
\newcommand{\eq}[1]{(\ref{#1})}

\newcommand{\halft}{{\textstyle \frac{1}{2}}}
\newcommand{\beq}{\begin{equation}}
\newcommand{\eeq}{\end{equation}}
\newcommand{\nn}{\nonumber}
\newcommand{\beqa}{\begin{eqnarray}}
\newcommand{\eeqa}{\end{eqnarray}}
\newcommand{\be}{\begin{eqnarray}}
\newcommand{\ee}{\end{eqnarray}}
\newcommand{\beqat}{\begin{eqnarray*}}
\newcommand{\eeqat}{\end{eqnarray*}}
\newcommand{\ket}[1]{\vert{#1}\rangle}
\newcommand{\bra}[1]{\langle{#1}\vert}

\newcommand{\PL}[3]{Phys.~Lett.~{\bf {#1}},~{#2}~({#3})}
\newcommand{\NP}[3]{Nucl.~Phys.~{\bf {#1}},~{#2}~({#3})}
\newcommand{\PRD}[3]{Phys.~Rev.~{\bf D{#1}},~{#2}~({#3})}
\newcommand{\PRL}[3]{Phys.~Rev.~Lett.~{\bf {#1}},~{#2}~({#3})}

\newcommand{\PRe}[3]{Phys.~Rep.~{\bf {#1}},~{#2}~({#3})}

\newcommand{\inv}[1]{\frac{1}{#1}}

\begin{document}

\title{Comments on the Relativity of Shape\footnote{Invited talk at the Workshop on the Shape of Hadrons, 27-29 April 2006 in Athens, Greece.}}

\author{Paul Hoyer}{
address={Department of Physical Sciences and Helsinki Institute of
              Physics\\
              POB 64, FIN-00014 University of Helsinki, Finland\\ {\rm and}},
,altaddress={NORDITA, Blegdamsvej 17, DK-2100 Copenhagen, Denmark}}

\begin{flushright}
\today\\ HIP-2006-34/TH \\ NORDITA-2006-24 \\ hep-ph/{0608295}
\end{flushright}

\begin{abstract}
In this talk I address three topics related to the shape of hadrons:
\begin{enumerate}
\item The Lorentz contraction of bound states. Few dedicated studies of this exist -- I describe a recent calculation for ordinary atoms (positronium).
\item Does the $A$-dependence of nuclear structure functions indicate a change of proton shape in the nuclear environment? (My short answer is no.)
\item The size of Fock states contributing to processes involving large momentum transfers. End-point configurations can be transversally extended and yet sufficiently short-lived to contribute coherently to hard scattering. 
\end{enumerate}
\end{abstract}
\maketitle

\section{Lorentz contraction of bound states}

We are all familiar with the classical prediction of special relativity that rigid rods appear contracted to a moving observer. Lorentz contraction is also often appealed to in descriptions of the scattering of hadrons and nuclei, which are depicted as thin disks when moving with high momenta. It is thus surprising that few studies seem to have been devoted to the contraction of bound state wave functions in relativistic quantum physics. I am aware of only a few examples \cite{Brodsky}, most of which are in the context of 1+1 dimensional models. Why do we not, in courses on relativistic quantum mechanics, derive the contracted wave function of the ordinary hydrogen atom in relativistic CM motion? What do we in fact know about the Fock states of the atomic wave function (even at lowest order in $\alpha$) when the (kinetic) energy of the atom is much larger than the masses of its constituents? Here I shall briefly describe some results of a recent study by my student Matti J\"arvinen \cite{Jarvinen:2004pi} of these issues, which are quite fascinating and deserve more attention.

\subsection{Wave functions and Fock states}

The wave function of a bound state can be defined in several ways, depending on how the time slice (snap-shot) is taken of the propagating state. In non-relativistic quantum mechanics we are used to {\it equal time} wave functions $\Psi_0(t;\xvec_1,\ldots,\xvec_N)$, which are probability amplitudes for finding the $N$ constituents at positions $\xvec_1,\ldots,\xvec_N$ at a common time $t$ (see Fig.~1a). The concept of equal time is frame dependent, which makes boosting $\Psi_0$ wave functions non-trivial.

\begin{figure}[h]
\resizebox{.7\textwidth}{!}
{\includegraphics{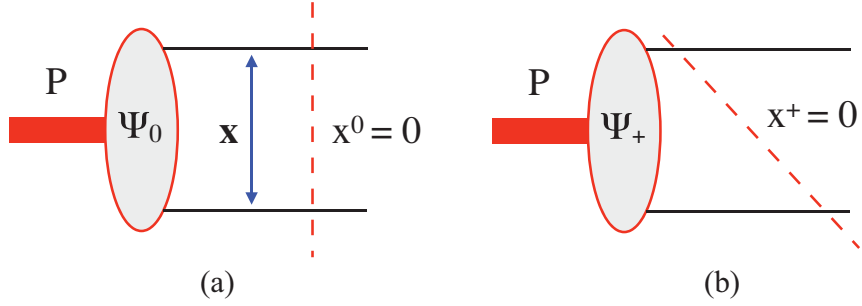}}
\caption{(a) The equal time wave function $\Psi_0$ is the amplitude for finding the constituents of a bound state at relative distance $\xvec$ when measured at a common time $t$. (b) The light-front wave function $\Psi_+$ is the amplitude for finding the constituents of a bound state at relative distance $(x^-,\xvec_\perp)$ when measured at a common LF time $x^+=t+z$.}
\label{fig1}
\end{figure}

In relativistic descriptions wave functions $\Psi_+(x^+;x_1^-,\xvec_{1\perp},\ldots,x_N^-,\xvec_{N\perp})$ defined at {\it equal light front (LF) time} $x^+ \equiv t+z$ are commonly used \cite{Brodsky:1997de} (Fig.~1b). Each constituent is then observed at a common $x^+ = x_1^+ = \ldots = x_N^+$, and the $\Psi_+$ wave function describes their distribution in $x^- \equiv t-z$ and $\xvec_\perp = (x,y)$. This LF wave function is {\it invariant under boosts} along the $z$-axis. Such boosts leave the (relative) LF time invariant: ${x'}^+ = e^\zeta x^+$, so that $x^+=0$ transforms into ${x'}^+=0$ for any boost parameter $\zeta$.

When the constituents move non-relativistically it makes no difference whether they are observed at equal $t$ or equal $x^+$: $\Psi_0=\Psi_+$ since the constituents move only an infinitesimal distance in the time it takes a ray of light to connect them (Fig.~1b). However, when an internally non-relativistic system like the Hydrogen atom is in relativistic motion also its constituents move relativistically. Then the two definitions of the wave function are no longer equivalent. Which one should be expected to exhibit Lorentz contraction?

\begin{figure}[h]
\resizebox{.3\textwidth}{!}
{\includegraphics{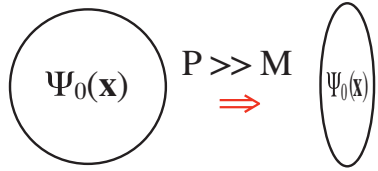}}
\caption{Lorentz contraction for the equal time wave function $\Psi_0(t,\xvec)$, as it is commonly depicted.}
\label{fig2}
\end{figure}
In the classical derivation of Lorentz contraction each observer measures the length of the rod {\it when its ends are at the same (ordinary) time $t$}. Contraction is the result of the different perception of time for observers in relative motion. It is thus clear that we should expect contraction for the equal (ordinary) time wave function $\Psi_0$ (Fig.~2). As we already noted, $\Psi_+$ is in fact invariant under boosts.

The pictorial representation of contraction in Fig.~2 leaves, however, many open questions. A field theory bound state is defined as the solution of the Schr\"odinger equation
\beq \label{Seq}
{\rm\hat H}\ket{N(\bs{P},s)} = E(\bs{P})\ket{N(\bs{P},s)}
\eeq
where the hamiltonian ${\rm \hat H}$ contains operators that create and annihilate particles\footnote{Since we are dealing with motion where the kinetic energies exceed particle masses we cannot restrict ourselves to Fock states with a fixed particle number. Even for the Hydrogen atom, Fock states beyond $e^+e^-$ can be neglected only in the rest frame (and at lowest order in $\alpha$).}. The bound state is a superposition of Fock states, \eg, for the proton
\beq
\ket{N} = \int \left[\prod_i \frac{dx_i\,d^2\kvec_{\perp i}}{16\pi^3}\right] \Big[\psi_{uud}(x_i,\kvec_{\perp i},\lambda_i)\ket{uud}
+ \psi_{uudg}(\ldots)\ket{uudg}+ \ldots \psi_{\ldots}\ket{uudq\bar q} + \ldots\Big]
\eeq
Should all the various Fock components contract at a universal rate? This would imply a simple relation between equal-time wave functions in different frames -- which is not expected since boosts are dynamical operators \cite{Dirac}. Boosting a rest frame wave function is in fact as complicated as solving the Schr\"odinger equation \eq{Seq} directly in the boosted frame. 

\begin{figure}[h]
\resizebox{.4\textwidth}{!}
{\includegraphics{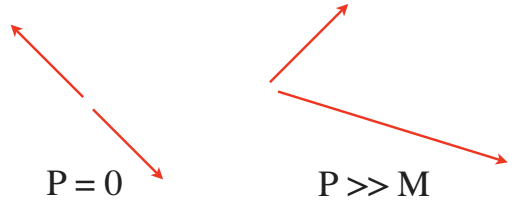}}
\caption{Momenta which are equal and opposite in the rest frame can be boosted into the forward hemisphere.}
\label{fig3}
\end{figure}
A Lorentz contraction changes the magnitudes of longitudinal momenta, leaving the transverse momenta unchanged. In particular, contractions do not change the sign of longitudinal momenta.  Boosts on the other hand can readily turn backward moving particles into forward moving ones, as indicated in Fig.~3. Thus it would be miraculous if bound state wave functions only would contract under boosts.

\begin{figure}[h]
\resizebox{.6\textwidth}{!}
{\includegraphics{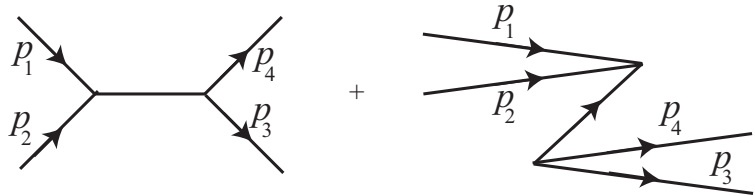}}
\caption{In old-fashioned perturbation theory there is a separate diagram for each time ordering of the interaction vertices. Only the sum of all diagrams is Lorentz invariant.}
\label{fig4}
\end{figure}
It is also well-known that Fock states mix under boosts. A simple $\phi^3$ example is shown in Fig.~4 using time-ordered (old-fashioned) perturbation theory. The two diagrams differ by the time ordering of their interaction vertices. The first diagram of Fig.~4 has a single particle in the intermediate state, while the second one has five. Their relative magnitude is frame dependent, but their sum is Lorentz invariant,
\beq
\frac{g^2}{2E}\left[\inv{E_1+E_2-E}-\inv{E_1+E_2+E}\right] = \frac{g^2}{(p_1+p_2)^2-M^2}
\eeq
where $E_i=\sqrt{\pvec_i^2+m^2}$ and $E=\sqrt{(\pvec_1+\pvec_2)^2+M^2}$. In the limit of very large boosts (\ie, in the infinite-momentum frame) only the first diagram survives -- and the rules are then the same as those of LF time ordered perturbation theory.

\subsection{Positronium in relativistic motion}

A relativistic bound state has (infinitely) many Fock components, making it practically impossible to derive the equal time wave function in the rest system, and even less so in a general frame. However, the Hydrogen atom (or positronium, for simplicity) has a very simple wave function in the rest frame. At lowest order, only the $\ket{e^+e^-}$ Fock state of positronium contributes, bound by the instantaneous Coulomb potential. The internal motion is non-relativistic, suggesting essential simplifications also in a frame where the bound state moves with relativistic speed. Thus we may ask whether the wave function of the $\ket{e^+e^-}$ component is Lorentz contracted? Do higher Fock states contribute, and do their wave functions also contract?


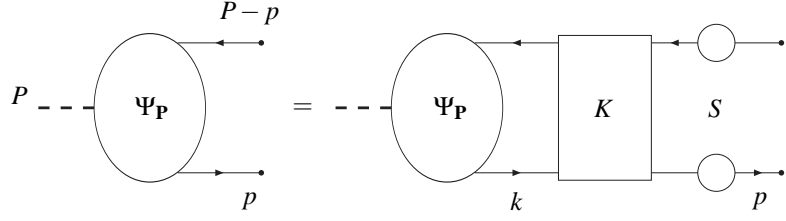
\begin{figure}[h]
\centering
 \SetScale{0.7}\begin{picture}(300,80)(-33,0)
  \Oval(70,50)(40,30)(0)
  \SetWidth{1.5}
  \DashLine(10,50)(40,50){7}
  \SetWidth{0.5}
  \ArrowLine(130,85)(85,85)
  \ArrowLine(85,15)(130,15)
  \Vertex(130,15){1.5}
  \Vertex(130,85){1.5}
  \Text(107,35)[]{$=$}
  \Oval(230,50)(40,30)(0)
  \SetWidth{1.5}
  \DashLine(170,50)(200,50){7}
  \SetWidth{0.5}
  \ArrowLine(290,85)(245,85)
  \ArrowLine(245,15)(290,15)
  \Boxc(315,50)(50,78)
  \ArrowLine(365,85)(340,85)
  \Oval(375,85)(10,10)(0)
  \Line(385,85)(410,85)
  \ArrowLine(385,15)(410,15)
  \Oval(375,15)(10,10)(0)
  \Line(340,15)(365,15)
  \Vertex(410,15){1.5}
  \Vertex(410,85){1.5}
  \Text(221,35)[]{$K$}
  \Text(263,35)[]{$S$}
  \Text(50,35)[]{$\Psi_{{\bf P}}$}
  \Text(163,35)[]{$\Psi_{{\bf P}}$}
  \Text(0,40)[]{$P$}
  \Text(87,0)[]{$p$}
  \Text(87,71)[]{$P-p$}
  \Text(280,0)[]{$p$}
  \Text(188,0)[]{$k$}
 \end{picture}
 \caption{The Bethe-Salpeter equation. The blobs represent the positronium wave function $\Psi_{{\bf P}}$, $K$ is the interaction kernel and $S$ is the $e^+e^-$ propagator.}
 \label{waveeq}
\end{figure}

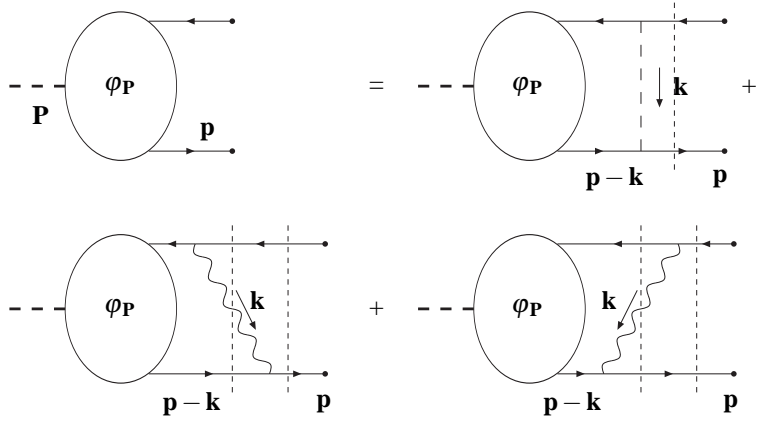
\begin{figure}[h]
\centering
\SetScale{0.7}\begin{picture}(300,155)(20,-85)
 \Oval(90,50)(40,30)(0)
 \SetWidth{1.5}
 \DashLine(30,50)(60,50){7}
 \SetWidth{0.5}
 \ArrowLine(150,85)(105,85)
 \ArrowLine(105,15)(150,15)
 \Vertex(150,15){1.5}
 \Vertex(150,85){1.5}
 \Text(63,35)[]{$\varphi_{\,\Pvec}$}

 \Text(160,35)[]{=}

 \Oval(310,50)(40,30)(0)
 \SetWidth{1.5}
 \DashLine(250,50)(280,50){7}
 \SetWidth{0.5}
 \ArrowLine(370,85)(325,85)
 \ArrowLine(325,15)(370,15)
 \DashLine(370,15)(370,85){7}
 \ArrowLine(370,15)(415,15)
 \ArrowLine(415,85)(370,85)
 \Vertex(415,85){1.5}
 \Vertex(415,15){1.5}
 \DashLine(388,5)(388,95){3}
 \Text(33,24)[]{$\Pvec$}
 \Text(96,18)[]{$\pvec$}
 \LongArrow(380,60)(380,40)
 \Text(275,35)[]{$\kvec$}
 \Text(290,0)[]{$\pvec$}
 \Text(250,0)[]{$\pvec-\kvec$}
 \Text(217,35)[]{$\varphi_{\,\Pvec}$}

 \Text(301,35)[]{+}

 \Oval(90,-70)(40,30)(0)
 \SetWidth{1.5}
 \DashLine(30,-70)(60,-70){7}
 \SetWidth{0.5}
 \ArrowLine(130,-35)(105,-35)
 \ArrowLine(105,-105)(170,-105)
 \Photon(130,-35)(170,-105){-3}{5}
 \ArrowLine(200,-35)(130,-35)
 \ArrowLine(170,-105)(200,-105)
 \Vertex(200,-35){1.5}
 \Vertex(200,-105){1.5}
 \DashLine(150,-25)(150,-115){3}
 \DashLine(180,-25)(180,-115){3}
 \Text(115,-46)[]{$\kvec$}
 \Text(140,-85)[]{$\pvec$}
 \Text(90,-85)[]{$\pvec-\kvec$}
 \LongArrow(152,-60)(162,-80)
 \Text(63,-49)[]{$\varphi_{\,\Pvec}$}

 \Text(160,-49)[]{+}

 \Oval(310,-70)(40,30)(0)
 \SetWidth{1.5}
 \DashLine(250,-70)(280,-70){7}
 \SetWidth{0.5}
 \ArrowLine(390,-35)(325,-35)
 \ArrowLine(325,-105)(350,-105)
 \Photon(390,-35)(350,-105){3}{5}
 \ArrowLine(420,-35)(390,-35)
 \ArrowLine(350,-105)(420,-105)
 \Vertex(420,-35){1.5}
 \Vertex(420,-105){1.5}
 \DashLine(370,-25)(370,-115){3}
 \DashLine(400,-25)(400,-115){3}
 \Text(248,-46)[]{$\kvec$}
 \Text(294,-85)[]{$\pvec$}
 \Text(234,-85)[]{$\pvec-\kvec$}
 \LongArrow(368,-60)(358,-80)
 \Text(217,-49)[]{$\varphi_{\,\Pvec}$}

 \end{picture}
\caption{The time-ordered bound state equation for positronium in the ladder approximation. The blobs denote the equal time wave function $\varphi_{\,\Pvec}$ of the $e^+e^-$ Fock component carrying total 3-momentum {\bf P}.}
\label{tobse}
\end{figure}

I shall here only summarize the results obtained in \cite{Jarvinen:2004pi}. It is necessary to start from the fully relativistic framework provided by the Bethe-Salpeter (BS) equation (Fig.~\ref{waveeq}). By inspecting the equation one may surmise that many contributions, such as electron pair production and annihilation, are of higher order in $\alpha$ and thus can be neglected in all frames. The relevant terms remaining in the interaction kernel are instantaneous photon exchange (as in the rest frame) and transverse photon exchange. Thus at lowest order the BS equation simplifies to that shown in Fig.~\ref{tobse}. As might have been expected, the boost has given the Coulomb exchange of the rest frame a transverse, propagating component. The propagation time is of the order of the Bohr radius, $1/(\alpha m_e)$. Since the binding energy of positronium $\Delta E \propto \alpha^2 m_e$, the lifetime of the $\ket{e^+e^-}$ Fock state is longer by a factor $1/\alpha$ compared to the lifetime of the $\ket{e^+e^-\gamma}$ state. The likelihood of ``seeing'' a photon in flight is thus of \order{\alpha}, even though transverse photon exchange contributes to positronium binding at leading order. The probability of catching {\em two} photons in flight is of \order{\alpha^2} and can be neglected.


The leading order BS equation shown in Fig.~\ref{tobse} can be readily solved. Due to the time ordering explicit Lorentz symmetry is lost. Hence there is no guarantee that the positronium energy is correctly given as $E=\sqrt{\Pvec^2+M^2}$, where $M$ is the rest mass -- except that we know that QED is Poincar\'e invariant. Indeed, this dependence of the energy on the CM momentum $\Pvec$ emerges almost miraculously from the equations. The $\ket{e^+e^-}$ Fock amplitude $\varphi_{\Pvec}$ satisfies
\beq
\left[E_b-\inv{m_e}\left(\pvec_\perp^2+\gamma^{-2}p_\parallel^2\right)\right]\,\varphi_{\,\Pvec}(\pvec) = - \frac{4\pi\alpha}{\gamma}\int\frac{d^3\kvec}{(2\pi)^3} \frac{\varphi_{\,\Pvec}(\pvec-\kvec)}{\kvec_\perp^2+\gamma^{-2}k_\parallel^2}
\eeq
where $E_b \equiv \sqrt{E^2-\Pvec^2}-2m_e$, $\gamma \equiv \sqrt{\Pvec^2+(2m_e)^2}/2m_e$ and $\pvec$ is the relative momentum between the electron and the positron. The scaling $p_\parallel \to \gamma p_\parallel$ and $k_\parallel \to \gamma k_\parallel$ eliminates all explicit $\Pvec$-dependence from the bound state equation. This means that $E_b$ is independent of $\Pvec$ as it should be, and that $\varphi_{\,\Pvec}$ Lorentz contracts.

The wave function of the $\ket{e^+e^-\gamma}$ Fock component can be similarly found. The probability distribution of the photon (integrated over the electron momenta) turns out to factorize for the positronium ground state,
\beq\label{photdist}
 \frac{d^2 \mathcal P}{d k_c\, d\! \cos \theta_c} = \frac{\alpha}{2 \pi} f(\cos\theta_c) g(k_c)
\eeq 
where $\kvec_c =(\kvec_\perp,k_\parallel/\gamma)$ is the photon momentum corrected for contraction and $\theta_c$ is the angle between $\Pvec$ and $\kvec_c$. The distributions are
\beqa\label{fdef}
 f(\cos\theta_c) &=& \frac{\gamma\beta^2 (1-\cos^2 \theta_c)}{(1+\beta^2\gamma^2\cos^2\theta_c)^{3/2}\left(\sqrt{1+\beta^2\gamma^2\cos^2\theta_c}-\beta \gamma \cos \theta_c \right)^2} \\ \nn \\
 g(k_c) &=& \frac{1}{k_c}\int \frac{d^3 \qvec_c}{(2\pi)^3} \left|\varphi_{0}(\qvec_c+\kvec_c/2)-\varphi_{0}(\qvec_c- \kvec_c/2)\right|^2
\eeqa
where $\varphi_{\,0}$ is the $\ket{e^+e^-}$ Fock amplitude in the rest frame ($\Pvec=0$), and $\beta=\sqrt{1-1/\gamma^2}$. The angular distribution \eq{fdef} of the photon depends on the CM motion ($\beta, \gamma$) even after rescaling. Thus this Fock component does not simply Lorentz contract, as we already anticipated (\cf\ Fig.~3). As shown in Fig.~7, the photon distribution changes with increasing $\beta$ from being symmetric (for $\beta \simeq 0$) into one that for $\beta\to 1$ has support only in the forward hemisphere ($\cos\theta_c >0$). The latter (infinite momentum frame) distribution agrees with that of the wave function defined at equal LF time $x^+ = t+z$ (which is invariant under boosts).

\begin{figure}[t]
\resizebox{.7\textwidth}{!}
{\includegraphics{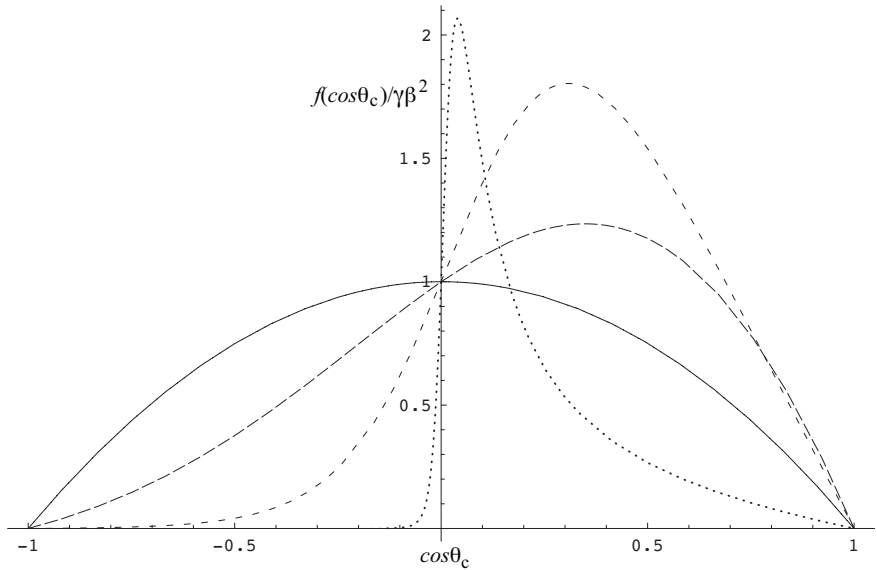}}
\caption{The angular dependence of the contracted and integrated photon distribution (\ref{photdist}) in the positronium ground state. The lines show the  angular distribution $f(\cos \theta)/(\gamma\beta^2)$ [defined in (\ref{fdef})] for $\beta=0.001$, $0.5$, $0.9$ and $0.999$. For $\beta=0.001$ (solid line) the distribution is close to the symmetric limit of the rest frame. For $\beta=0.999$ (dotted line) the distribution approaches the infinite momentum frame (alias LF wave function) where the momenta of all particles are in the forward hemisphere.}
\label{angplot}
\end{figure}

The study \cite{Jarvinen:2004pi} of the frame dependence of the Hydrogen (or positronium) equal time wave function has thus revealed a richness beyond the standard picture of Fig.~2. The wave function of the lowest $\ket{e^+e^-}$ Fock state (which is non-relativistic in the rest frame) indeed Lorentz contracts. Note, however, that this contraction is in the {\em relative} longitudinal momentum of the constituents. In a positronium state of momentum $\Pvec$ both the electron and the positron momenta equal $\Pvec/2$, up to the small relative momentum of \order{\alpha}. For $\Pvec=0$ (the left hand side of Fig.~2) the momenta of the constituents are in opposite hemispheres. For $\Pvec$ larger than the Bohr momentum both the electron and the positron momenta are in the forward hemisphere, the flattened disk on the right hand side of Fig.~2 describing only their deviations from $\Pvec/2$.

For the photon in the $\ket{e^+e^-\gamma}$ Fock state the situation is different since it carries only a fraction \order{\alpha} of the bound state momentum $\Pvec$. As seen in Fig.~7, the photon momentum can be in the backward hemisphere even for large $\Pvec$, being restricted to the forward hemisphere only in the infinite momentum frame ($\Pvec\to\infty$).

Further information on frame dependence could be obtained from studies of the higher order corrections to the positronium wave function including, \eg, Fock states with an additional fermion pair. It should also be possible to study the CM momentum dependence of non-relativistic bound states in theories with other (\eg, $\phi^3$) interactions. The frame dependence of a state such as the proton, which is relativistic also in its rest frame, is likely to be more involved than that of positronium. It is not obvious that {\em any} Fock component of the proton (including the lowest one, $\ket{uud}$) obeys the classical longitudinal contraction sketched in Fig.~2.

\section{Shape of Nucleons in Nuclei}

The discovery of a nuclear ($A$-)dependence of parton distributions raised the question whether the cause is to be found in an actual modification of the nucleon wave function in the nuclear environment, or reflects the dynamics of the Deep Inelastic Scattering (DIS) measurement \cite{Arneodo}. To some extent this is a matter of definition. The parton distribution shown in Fig.~8,
\begin{figure}[t]
\resizebox{.6\textwidth}{!}
{\includegraphics{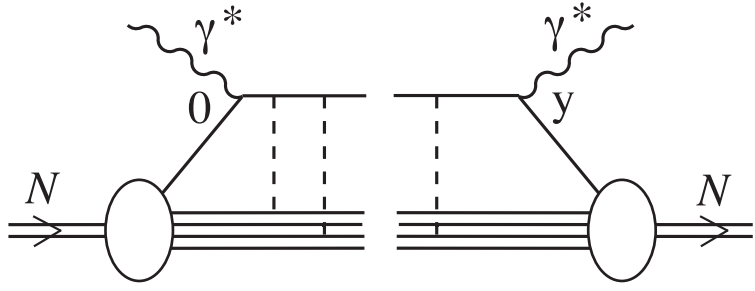}}
\caption{The parton distribution \eq{partdist}. Rescattering arising from the Wilson line \eq{Wline} is shown as dashed lines.}
\label{Fig8}
\end{figure}

\beq\label{partdist}
f_{q/N}(x_B,Q^2)= \inv{8\pi}\int dx^- e^{-ix_Bp^+x^-/2} \bra{N(p)}\bar q(x^-)\gamma^+ W[x^-,0]q(0)\ket{N(p)}_{x^+=0}
\eeq
(where $x^\pm=t\pm z$) involves the Wilson line
\beq\label{Wline}
W[x^-,0] \equiv {\rm P} \exp\left[\frac{ig}{2}\int_0^{x^-}dw^- A^+(w^-)\right]
\eeq
which may be included in the definition of the LF wave function (which then becomes specific to DIS and similar hard processes). On the other hand, the Wilson line may also be regarded as reflecting physical rescattering of the struck quark on target spectators \cite{Brodsky:2002ue} (dashed lines in Fig.~8). Since there are more spectators in a nuclear environment the rescattering can be expected to be $A$-dependent. This view of rescattering is the same as the one usually formulated in the ``target rest frame'' in models of shadowing \cite{Piller:1999wx,Armesto:2006ph}. In either case, the rescattering must be coherent with the hard virtual photon interaction in order to affect the total DIS cross section. This coherence allows the different physical interpretations to be equivalent.

\begin{figure}[h]
\resizebox{.5\textwidth}{!}
{\includegraphics{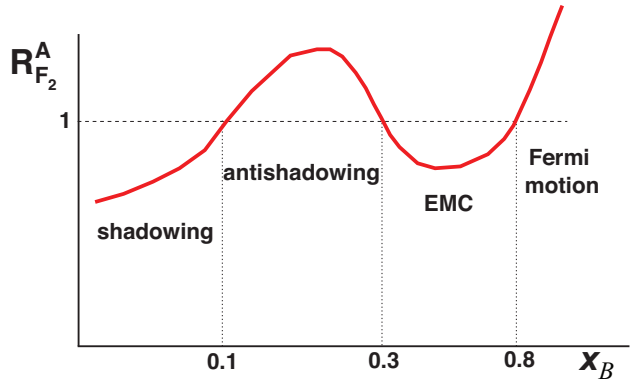}}
\caption{Schematic behaviour of the ratio of the nuclear to free nucleon $F_2$ structure functions $R_{F_2}^A(x_B,Q^2)$ as function of Bjorken $x_B$ for a given fixed $Q^2$. From Ref.~\cite{Armesto:2006ph}.}
\label{Fig9}
\end{figure}
\begin{figure}[h]
\resizebox{.3\textwidth}{!}
{\includegraphics{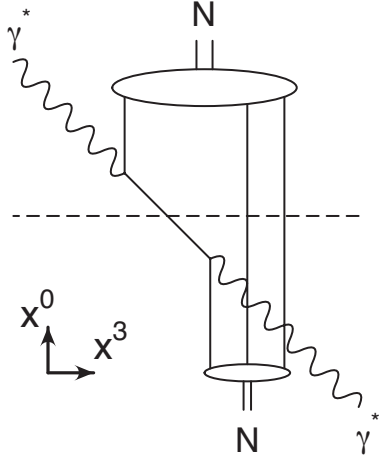}}
\caption{The matrix element in the integrand of the parton distribution \eq{partdist}, \ie, the handbag diagram of Fig.~8 viewed in coordinate space (rescattering is not shown). The position of the struck quark differs by $x^-$ in the two wave functions (whereas $x^+=\xvec_\perp=0$).}
\label{Fig10}
\end{figure}

As sketched in Fig.~9, the $A$-dependence extends beyond the shadowing region of small $x_B$. One may ask whether the ``antishadowing'' and ``EMC'' effects reflect an actual change of the nucleon shape, beyond effects due to rescattering of the DIS probe. One answer is provided by the observation made several years ago \cite{Hoyer:1996nr} that these structures essentially vanish if one views the parton distribution in coordinate ($x^-$) space, rather than in momentum ($x_B$) space. The parton distribution \eq{partdist} is a Fourier transform of an overlap of nucleon wave functions, with a shift $x^-$ in the position of the struck quark (Fig.~10). We may express the the matrix element in terms of the parton distributions through the inverse Fourier transform,
\beq
\bra{N}\bar q(x^-)\gamma^+ W[x^-,0]q(0)\ket{N}-(x^- \to -x^-)= 4im\int_0^1 dx_B \big[f_{q/N}(x_B)+f_{\bar q/N}(x_B)\big] \sin\left(\halft mx_B x^- \right)
\eeq
The \rhs of this equation is essentially given by the $F_2$ structure function. Thus we can study the $A$-dependence of the parton distribution in coordinate space, defined as
\beq\label{partcoord}
q^A(x^-,Q^2) \equiv \int_0^1 \frac{dx_B}{x_B} F_2^D(x_B,Q^2) R_{F_2}^A(x_B,Q^2) \sin\left(\halft mx_B x^-\right)
\eeq
where $R_{F_2}^A(x_B,Q^2)$ is the experimentally measured ratio of nuclear to deuterium structure functions sketched in Fig.~9. The corresponding ratio in coordinate space, defined as
\beq\label{ratiocoord}
R^A(x^-,Q^2) \equiv \frac{q^A(x^-,Q^2)}{q^D(x^-,Q^2)}
\eeq
can then be formed using data on structure functions and is shown in Fig.~11a. 
\vspace{.3cm}
\begin{figure}[h]
\resizebox{.7\textwidth}{!}
{\includegraphics{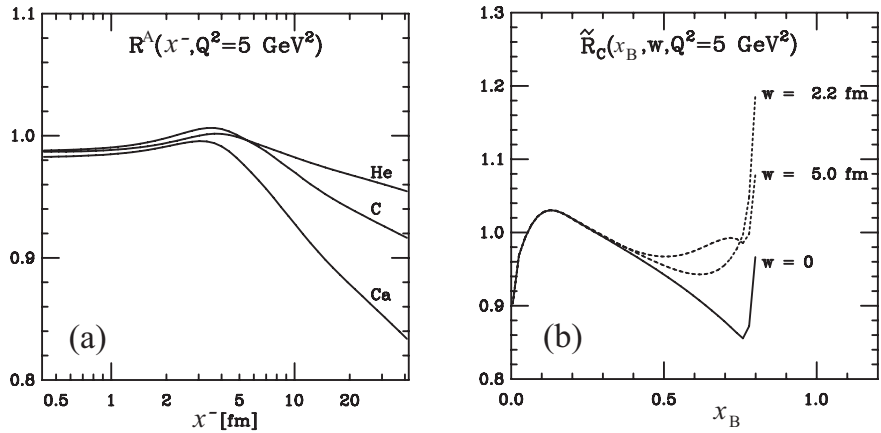}}
\caption{(a) The coordinate space ratio $R^A(x^-,Q^2)$ \eq{ratiocoord} obtained by Fourier transforming data on $F_2^A(x_B,Q^2)$ structure functions for $A=$ He, C and Ca. (b) The momentum space ratio $\tilde R^C(x_B,w,Q^2=5\ {\rm GeV}^2)$ for Carbon, obtained by Fourier transforming a modified coordinate space distribution in which all nuclear effects are eliminated for $x^- < w$.}
\label{Fig11}
\end{figure}

Within the {\it ca.} 1\% error bars \cite{Hoyer:1996nr} the ratio $R^A(x^-,Q^2)$ is consistent with having {\em no $A$-dependence} for $x^- \lsim 5$ fm. At longer distances $x^- > 5$ (\ie, $t=-z>2.5$ fm since $x^+=0$) shadowing sets in. Thus viewed from coordinate space (which is not unnatural for discussing effects of nuclear size) we may regard\footnote{Understanding the dynamics of nuclear dependence in momentum space is nevertheless interesting in its own right. See \cite{Brodsky:2004qa} for recent ideas about the origin of the antishadowing enhancement.} antishadowing and the EMC effect as merely resulting from Fourier transforming a flat distribution (of finite length) in $x^-$! This is corroborated in Fig.~11b, where the reverse transform back to momentum ($x_B$-) space is made, under the assumption that $R^A(x^-,Q^2)$ is unity for $x^-< w$ (and takes the values of Fig.~11a for $x^- > w$). It is seen that the antishadowing and (most of) the EMC effect is reproduced assuming no nuclear dependence in coordinate space for $x^-\lsim 5$ fm. The nuclear effects can thus be ascribed solely to shadowing.

The parton distribution $q^A(x^-,Q^2)$ in coordinate space is insensitive to the region of Fermi motion at large $x_B$ in Fig.~9, where the structure function $F_2(x_B,Q^2)$ is small. 
The sizeable nuclear dependence of $R_{F_2}^A(x_B,Q^2)$ at large $x_B$ reflects the ratio of very small $F_2$, which do not appreciably affect the inverse Fourier transform \eq{partcoord}.

\section{Size of Hard Subprocesses}

The third aspect of shape that I would like to discuss concerns the size of coherent hard subprocesses in scattering involving large momentum transfers. As sketched in Fig.~12, in inclusive DIS ($ep\to eX$) we expect that the virtual photon (whose transverse coherence length is $\sim 1/Q$) scatters off a single quark. The quark is typically part of a Fock state with a hadronic, $\sim 1$ fm size. In elastic scattering ($ep \to ep$), where the entire Fock state must coherently absorb the momentum, one might on the other hand expect \cite{bl:1980} that only compact Fock states of the photon, with transverse sizes $r_\perp \sim 1/Q$ will contribute. Thus the dynamics of inclusive and exclusive processes appears to be quite different. In particular, the dependence on the electric charges of the quarks is expected to be, qualitatively,
\beqa
\sigma(ep \to eX) &\propto& \sum_q e_q^2 \hspace{2cm} {\rm Inclusive,\ DIS}\nn\\ \\ 
\sigma(ep\to ep) &\propto& (\sum_q e_q)^2  \hspace{1.5cm} {\rm Exclusive,\ form\ factor}\nn
\eeqa

\begin{figure}[h]
\resizebox{.9\textwidth}{!}
{\includegraphics{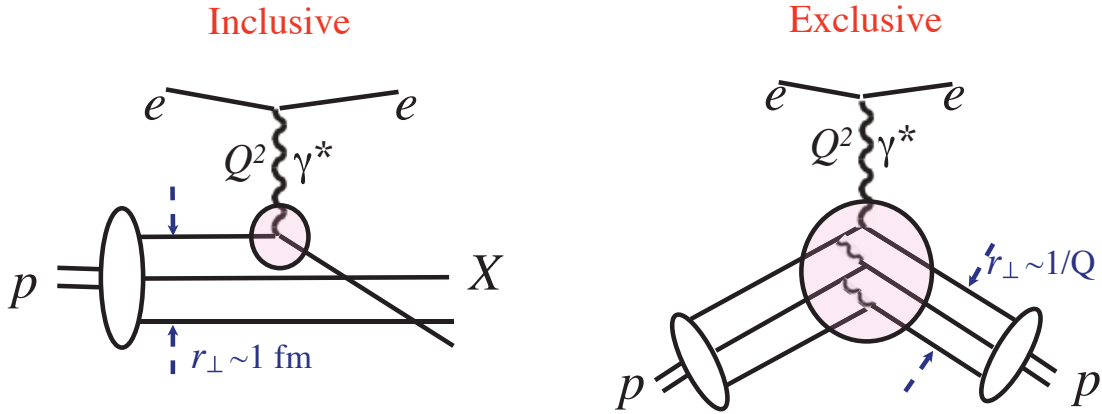}}
\caption{The virtual photon scatters from single quarks in inclusive deep inelastic scattering {\it (left)}. If the valence quarks absorb equal shares of the momentum transfer in the exclusive $ep \to ep$ process {\it (right)} only compact Fock states can contribute.}
\label{Fig12}
\end{figure}

In contrast to these expectations the data suggests a close connection between inclusive and exclusive scattering. The resonance production $ep \to eN^*$ cross sections (including $N^*=p$) average the DIS scaling curve when plotted at the same value of $x_B$ (or of the related Nachtmann variable $\xi$) \cite{Melnitchouk:2005zr}. Examples of this Bloom-Gilman duality are shown in Fig.~13. A natural explanation of duality is that the same Fock states of the proton contribute in both cases \cite{Hoyer:2005nk}. Resonance formation occurs on a longer time scale than the hard subprocess, hence is incoherent with it and cannot change the total cross section. Only the local mass distribution (resonance bumps) is sensitive to the hadronization time scale.

\begin{figure}[h]
\resizebox{1.0\textwidth}{!}
{\includegraphics{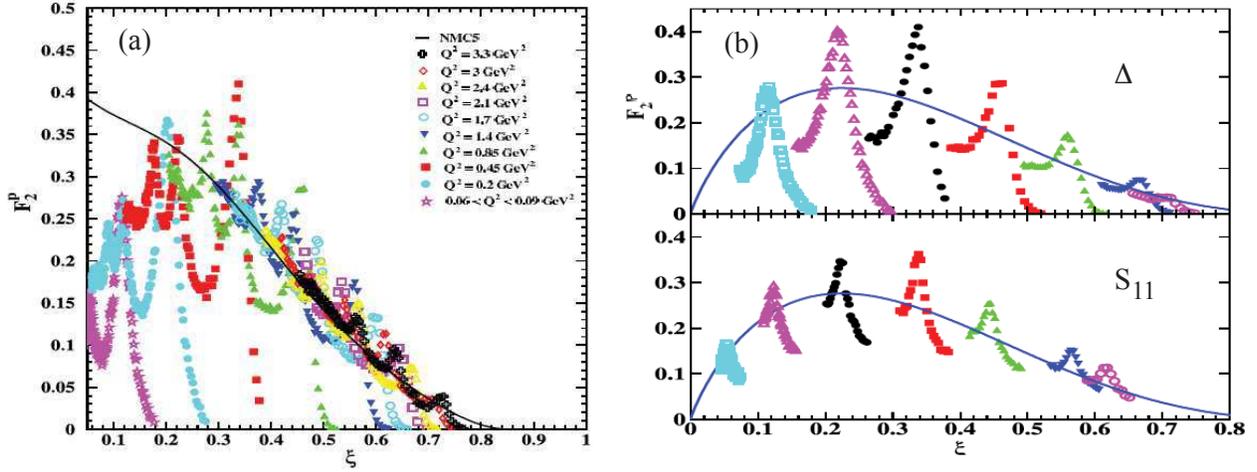}}
\caption{Bloom-Gilman duality. (a) Resonance production in exclusive $ep \to eN^*$ average the scaling curve measured at high $Q^2$ as a function of the Nachtmann variable $\xi$ (\ie, $x_B$ with a target mass correction). (b) Proton $F_2$ structure function in the $\Delta$ (top) and $S_{11}$ (bottom) resonance regions from Jefferson Lab Hall C, compared to the scaling curve from Ref. \cite{Niculescu:2000tk}. The resonances move to higher $\xi$ with increasing $Q^2$, which ranges from $\sim 0.5\ \gev^2$ (smallest $\xi$ values) to $\sim 4.5\ \gev^2$ (largest $\xi$ values). From Ref. \cite{Melnitchouk:2005zr}.}
\label{Fig13}
\end{figure}

As may be seen from the \rhs of Fig. 13 the resonance bumps migrate with increasing $Q^2$ to large $x_B$, where the struck quark carries most of the proton momentum. This is the ``end-point'' region where the Lepage-Brodsky picture \cite{bl:1980} of exclusive form factors breaks down \cite{ilr}. When the struck quark carries nearly all the proton momentum the remaining (soft) partons can join the wave function of the final proton without the hard gluon exchanges indicated on the \rhs of Fig.~12 (this is the `Feynman mechanism'). The soft partons have short formation times $\propto 1-x_B$ and can thus stay coherent with the hard process, even though they are broadly distributed ($r_\perp \sim 1$ fm) \cite{Brodsky:1991dj}.

The correct dynamics of exclusive form factors is still an open and important issue. Further hints may be obtained from spin dependence. Here I would like to mention the photoproduction process $\gamma p \to \rho p$, which has been measured\footnote{The final proton was not observed and may dissociate into a low mass system.} at large momentum transfer $|t| \lsim 10\ \gev^2$ at HERA \cite{Chekanov:2002,Aktas:2006qs}. Due to the high energy one expects two-gluon exchange to dominate. The upper vertex (Fig.~14) should be hard due to the large momentum transfer $|t|$ carried by the gluons. Consequently the quark helicity is conserved at the gluon and photon vertices and the sum of the quark and antiquark helicities vanish, as indicated in Fig.~14. In the absence of orbital angular momentum (which should be suppressed due to the small transverse size of the quark pair) the $\rho$ meson will then be longitudinal ($\lambda_\rho=0$). 

\begin{figure}[h]
\resizebox{.6\textwidth}{!}
{\includegraphics{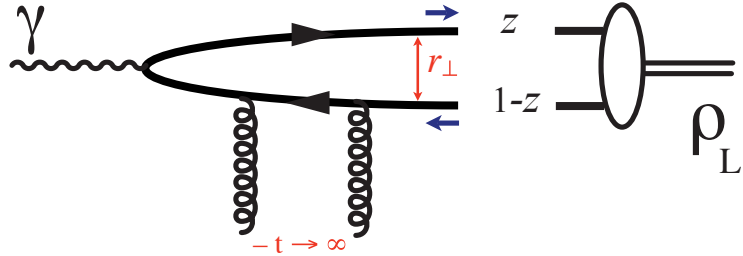}}
\caption{Upper vertex of the photoproduction process $ep \to \rho p$. The photon creates a quark and an antiquark which carry fractions $z$ and $1-z$ of its momentum. The pair scatters with large momentum transfer via two-gluon exchange and then forms the $\rho$ meson. The transverse size $r_\perp$ of the quark pair is of $\mathcal{O}(1/\sqrt{|t|})$ unless $z \to 0$ or $z \to 1$.}
\label{Fig14}
\end{figure}

The data shows that the $\rho$ is dominantly produced with helicity $\lambda_\rho=\pm 1$. The density matrix element 
\beq\label{spindense}
r_{00}^{04} \simeq \frac{T_{01}^2}{T_{01}^2+T_{11}^2+T_{-11}^2}
\eeq
where the $T_{\lambda_\rho\lambda_\gamma}$ are helicity amplitudes, is small in the whole range of $|t|\lsim 6\ \gev^2$ (Fig.~15). Thus quark helicity is maximally violated, while the $\gamma$ and $\rho$ helicities are approximately equal.

\begin{figure}[h]
\resizebox{.8\textwidth}{!}
{\includegraphics{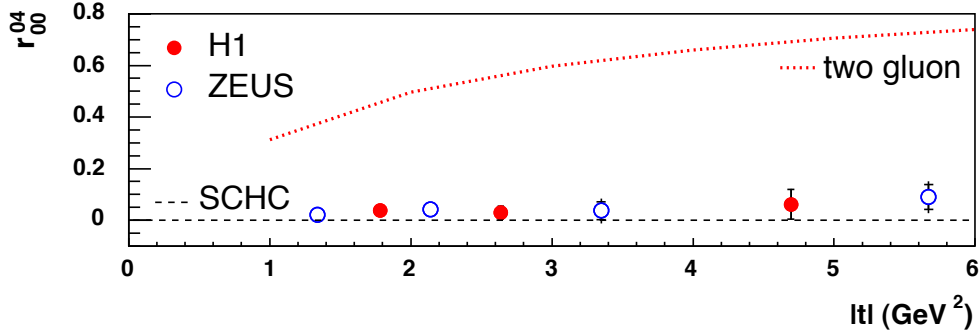}}
\caption{The spin density matrix element $r_{00}^{04}$ of \eq{spindense} for $\rho$ meson
photoproduction as a function of |t| from H1 \cite{Aktas:2006qs} (full points) and ZEUS \cite{Chekanov:2002} (open points). The dotted line shows the prediction of a two-gluon exchange model. The dashed line shows the expectation from $s$-channel helicity conservation (SCHC). From Ref. \cite{Aktas:2006qs}.}
\label{Fig15}
\end{figure}

The perturbative $\gamma + (gg) \to q\bar q$ with a quark helicity flip such that the helicity of the quark pair is $\lambda_{q\bar q}=1$ is \cite{Hoyer:2002qg}
\beq\label{transamp}
{\mathcal A}(\gamma + (gg) \to q\bar q) = -\frac{\sqrt{2}ee_q(4\pi\as)}{\sqrt{-3t}}\frac{\sqrt{m_q^2/(-t)}}{z^2(1-z)^2}\left[1+\morder{\frac{m_q^2}{t}}\right] \hspace{2cm} (\lambda_{q\bar q}=1)
\eeq
This amplitude should be convoluted with the distribution amplitude of the $\rho$ meson, which typically has the form
\beq\label{distamp}
\Phi_\rho(z) \propto z(1-z)
\eeq 
The integral over $z$ is then logarithmically singular at $z=0,1$, indicating the potential importance of endpoint contributions \cite{Hoyer:2002qg}. 

When a quark carries low longitudinal momentum its transverse size distribution grows rapidly (due to the absence of Lorentz time dilation). It may then scatter softly and flip its helicity without the $m_q/\sqrt{-t}$ penalty of \eq{transamp}. This dynamics is actually quite similar to that of ordinary DIS viewed in the target rest frame. The virtual photon splits into a quark and an antiquark, with the antiquark carrying finite momentum in the target rest frame (even though the photon energy $\nu \to \infty$). The soft scattering of the antiquark in the target ``releases'' the quark, which forms the current jet.
In the photoproduction process $\gamma p \to \rho p$ the distribution amplitude \eq{distamp} of the $\rho$ meson suppresses the endpoint region, but apparently not sufficiently. Thus the dominant subprocess may actually have transverse size $r_\perp \sim 1$ fm rather than $r_\perp \sim 1/\sqrt{-t}$.

\section{ACKNOWLEDGMENTS}

I am much indebted to my collaborators on the various topics discussed above, and grateful to the organizers of the the ``Shape of Hadrons'' workshop for their kind invitation. This research was supported in part by the Academy of Finland through grant 102046.

\end{document}